\documentclass[twocolumn,showpacs,prl,amsmath,amssymb,superscriptaddress]{revtex4}
\usepackage{graphicx}

\begin{document}

\title{Microphotoluminescence study of disorder in ferromagnetic (Cd,Mn)Te quantum
well.}

\author{W.~Ma\'{s}lana}
\affiliation{ Institute of Experimental Physics, Warsaw
University, 69 Hoza, 00-681 Warszawa, Poland.}
\affiliation{"Nanophysics and semiconductors" group, Laboratoire
de Spectrom\'{e}trie Physique, CNRS et Universit\'{e} Joseph
Fourier-Grenoble, B.P. 87, 38402 Saint Martin d'H\`{e}res Cedex,
France}

\author{P.~Kossacki}
\affiliation{ Institute of Experimental Physics, Warsaw
University, 69 Hoza, 00-681 Warszawa, Poland.}
\affiliation{"Nanophysics and semiconductors" group, Laboratoire
de Spectrom\'{e}trie Physique, CNRS et Universit\'{e} Joseph
Fourier-Grenoble, B.P. 87, 38402 Saint Martin d'H\`{e}res Cedex,
France}

\author{P.~P\l ochocka}
\affiliation{ Institute of Experimental Physics, Warsaw
University, 69 Hoza, 00-681 Warszawa, Poland.}

\author{A.~Golnik}
\affiliation{ Institute of Experimental Physics, Warsaw
University, 69 Hoza, 00-681 Warszawa, Poland.}

\author{ J.A.~Gaj}
 \affiliation{ Institute of Experimental Physics, Warsaw
University, 69 Hoza, 00-681 Warszawa, Poland.}

\author{D.~Ferrand}
\affiliation{"Nanophysics and semiconductors" group, Laboratoire
de Spectrom\'{e}trie Physique, CNRS et Universit\'{e} Joseph
Fourier-Grenoble, B.P. 87, 38402 Saint Martin d'H\`{e}res Cedex,
France}

\author{M.~Bertolini}
 \affiliation{"Nanophysics and semiconductors" group, Laboratoire
de Spectrom\'{e}trie Physique, CNRS et Universit\'{e} Joseph
Fourier-Grenoble, B.P. 87, 38402 Saint Martin d'H\`{e}res Cedex,
France}

\author{S.~Tatarenko}
 \affiliation{"Nanophysics and semiconductors" group, Laboratoire
de Spectrom\'{e}trie Physique, CNRS et Universit\'{e} Joseph
Fourier-Grenoble, B.P. 87, 38402 Saint Martin d'H\`{e}res Cedex,
France}

\author{J.~Cibert}
 \affiliation{ Laboratoire Louis N\'{e}el, CNRS, BP166, 38042
Grenoble cedex 9, France}

\date{\today}

\begin{abstract}
Microphotoluminescence mapping experiments were performed on a
modulation doped (Cd,Mn)Te quantum well exhibiting carrier induced
ferromagnetism. The zero field splitting that reveals the presence
of a spontaneous magnetization in the low-temperature phase, is
measured locally; its fluctuations are compared to those of the
spin content and of the carrier density, also measured
spectroscopically in the same run. We show that the fluctuations
of the carrier density are the main mechanism responsible for the
fluctuations of the spontaneous magnetization in the ferromagnetic
phase, while those of the Mn spin density have no detectable
effect at this scale of observation.
\end{abstract}

\pacs{ 71.35.Pq, 71.70.Gm, 75.50.Dd, 78.55.Et, 78.67.De, 85.75.-d}
\maketitle

 The realization of carrier-induced ferromagnetism \cite{1,2},
with possible applications in such rapidly developing areas as
"spintronics", has boosted the interest in the study of diluted
magnetic semiconductors (DMS) with a large density of carriers.
The experimental work has been mainly focused on (Ga,Mn)As,
resulting in a rapid advancement in the control of defects in
layers grown by molecular beam epitaxy (MBE), and the achievement
of relatively high values of the critical temperature $T_c$. Some
devices have already been implemented \cite{3}. From the
theoretical point of view, a very simple mean field model (Zener
model) appears to be surprisingly successful \cite{1,4}. One main
advantage of this model is that it uses parameters, which are
deduced from spectroscopic studies. However, it neglects disorder
and fluctuations.

Disorder in these materials has many sources. The solubility of Mn
in GaAs is low, so that layers have to be grown at low
temperature, resulting in a large amount of structural defects;
some of them (As antisites, Mn interstitials) can be eliminated at
least partially by a careful adjustment of the III/V ratio and
post-growth annealing, what leads to higher values of the
conductivity and of $T_c$ \cite{5,6,7}. However, disorder is
intrinsically present in doped DMS, and it is widely recognized
that this is a crucial issue for understanding carrier-induced
ferromagnetism in semiconductors. Thermal fluctuations have been
early incorporated in the mean field model by considering
elementary excitations \cite{8}. The random distribution of the
localized spins results in spin-glass like behaviours in undoped
DMS \cite{9}, an effect which should be even enhanced by the
oscillatory character of the RKKY interaction mediated by free
carriers. Electrical doping of semiconductors also involves a
random distribution of electrically active impurities. The Zener
model assumes the presence of free carriers, and the opposite
limit of strongly localized carriers (where magnetic polarons
form) has been considered \cite{10,11}. However, actual samples
are on both sides of the metal-insulator transition, and close to
it. More recent models aim at fully incorporating disorder
\cite{12,13,14,15}. They will help to understand, what must be
considered as a specific consequence of disorder, and where its
effects should be searched experimentally.

Experimental studies of disorder in doped DMS's are scarce. In
addition, several features of carrier induced ferromagnetism in
DMS's, which have been sometimes considered as due to disorder,
are already present in the Zener model. As a first example, if the
two subsystems (localized spins and free carriers) are treated
separately, one of them will saturate first; in the case of DMS's,
the carrier density is small and the carrier gas is fully
polarized while the magnetic ions spin system is far from
saturation; as a result, the magnetization follows a Brillouin
function, which increases when a magnetic field is applied at
constant temperature \cite{16}, or when the temperature is
decreased at constant (even zero) field; in the latter case, an
upward curvature is expected \cite{17} which in this case has
nothing to do with disorder. Another example is the threshold in
the dependence of $T_c$ on carrier density or spin density, which
is reproduced \cite{2,17,18} if the short-range spin-spin
interactions are introduced in the mean field model (in a
phenomenological way, but without adjustable parameter).

A direct effect of disorder is the reduced value of the
spontaneous magnetization. This is a very common observation in
ferromagnetic DMS's, which has been rapidly attributed to the
presence of uncorrelated spins \cite{1}. In (Ga,Mn)As, such spins
are expected in those parts of the material with smaller Mn
content, and hence smaller spin and carrier density. The effect is
particularly strong in samples, which are on the insulating side
of the metal insulator transition. The study of nitrogen- doped
(Zn,Mn)Te \cite{18} gives further insight into the mechanism,
since in that case Mn is an isoelectronic impurity and carriers
originate from the nitrogen acceptor. It was clearly observed that
the measured spontaneous magnetization closely matches the
prediction from the mean field model in metallic samples, while it
is significantly smaller in the insulating ones \cite{18}. This
suggests that electrostatic disorder plays the most significant
role (It should be kept in mind however that the presence of the
Mn makes the Mott critical density to increase: this effect is
suppressed by applying a magnetic field, which confirms the role
of spin fluctuations \cite{18}). Another confirmation comes from
the study of modulation-doped (Cd,Mn)Te quantum wells (QW). The
measured magnetization was found \cite{19, 20} to be proportional
to both the spin density $x_{eff}$ and the carrier density $p$, in
agreement with the mean field model - in spite of the low values
of $p$. This observation can be ascribed to the strong reduction
of electrostatic disorder, achieved by using remote doping in such
structures, as compared to (Zn,Mn)Te layers.

As the magnetization in a (Cd,Mn)Te QW is usually deduced from the
giant Zeeman splitting of the semiconductor's photoluminescence
(PL), local measurements can be performed using micro-PL. We
present here a study of the static fluctuations of the local,
spontaneous magnetization in p-doped (Cd,Mn)Te QWs, i.e., in a DMS
structure where ({\it i}) thermal fluctuations are minimized (the
highest critical temperature observed for such systems was 6.7K),
({\it ii}) electrostatic disorder is smaller than in thick layers
(thanks to remote doping), and ({\it iii}) magnetic disorder is
due to the random distribution of Mn atoms. We examine how these
fluctuations correlate to those of the spin density and of the
carrier density, also deduced from micro-PL spectra from the same
run.

Micro-PL was measured in a specially designed set-up with a
microscope lens immersed in the pumped helium bath of a
superconducting magnet \cite{21}, at temperatures from 4.2 K down
to 1.5 K. PL maps of a $32\times32$ micrometer wide square area
were recorded by placing a 10mm thick plane-parallel quartz plate
between the sample and the first lens of the optical system and
rotating the plate around two perpendicular axes. PL was excited
using a semiconductor diode laser (680 nm) focused to a spot of
diameter smaller than 0.5 micrometer, and recorded using a CCD
camera at the exit plane of a spectrograph.

The sample was a single, 8 nm wide, (Cd,Mn)Te QW embedded between
(Cd,Mg)Te barriers (27\% Mg), grown by MBE technique on a (001)
oriented, 4\% (Cd,Zn)Te substrate. The QW contained 4.65\% of Mn,
as determined from reflectivity spectra in magnetic fields up to
4.5 T, by fitting the Zeeman splitting with a modified Brillouin
function \cite{22} (not shown) with the parameter
$x_{eff}=2.74\%$. The QW was covered by a 25~nm thick cap layer,
thin enough so that a hole gas of density about $3\times 10^{11}
\mathrm{cm}^{-2}$ was generated by acceptor surface states
\cite{24}. The use of surface states as a source of holes -
instead of nitrogen doping - allows us to grow the sample at the
optimal temperature of $280^\circ$C, so that pseudo-smooth
interfaces are realized \cite{25}. The white light illumination in
such experiments efficiently depletes the QW from its hole gas
\cite{23}.

The general properties of PL from a p-doped (Cd,Mn)Te QW have been
described in detail in ref. \onlinecite{26}. We recall here the
minimum information needed to extract the local values of $p$ and
$x_{eff}$ and the spontaneous magnetization $M_0$ from micro-PL
data, as summarized schematically in the central part of Fig.1.

\begin{figure*}
\includegraphics[width=10cm]{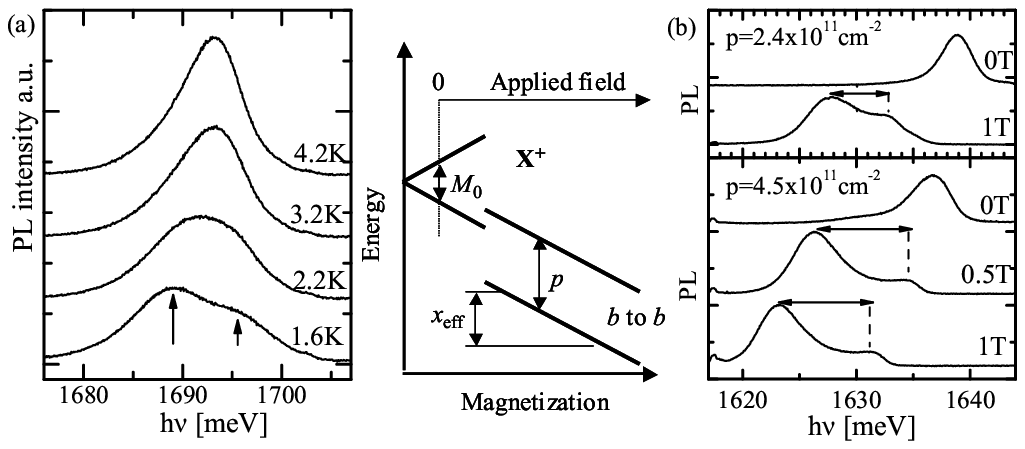}
\caption{Characteristic features of PL from a p-doped (Cd,Mn)Te
QW. (a) PL at zero field, from the sample with 4.65\% Mn, at
different temperatures. The arrows show the zero field splitting,
due to the giant Zeeman splitting of the charged exciton ($X^+$ in
the central scheme) induced by the spontaneous magnetization
$M_0$; (b) PL from a QW with 1.4\% Mn, at two values of the
carrier density (as deduced from the Moss-Burstein shift). The
double arrows on the double line (which is close to a band to band
transition, b to b in the central scheme) indicate the splitting
used to measure the carrier density.}
\end{figure*}

At small spin splitting (either spontaneous or field induced), PL
is due to the recombination of the positively charged exciton,
X$^+$. It takes place in $\sigma ^{+}$ polarization if the
recombined hole belongs to the majority spin subband, and in
$\sigma ^{-}$ for holes with minority spin (all being defined with
respect to the local magnetization). The splitting between the two
lines is the so-called giant Zeeman splitting; it is proportional
to the local magnetization. In a QW with high enough values of $p$
and $x_{eff}$, the onset of ferromagnetic ordering is witnessed by
a zero-field Zeeman splitting of the PL line, Fig. 1a \cite{2,17},
which signs the presence of a spontaneous magnetization inside the
QW, at temperatures below $T_c$. For the present sample, we
determined $T_c = 2.5$~K. In the paramagnetic phase, one observes
a critical divergence of the field-induced PL splitting when
decreasing the temperature, which follows a critical law, $\chi
(T)=C / (T-T_{CW})$ . No such effects are detected in undoped
structures or under white-light illumination (which significantly
reduces the hole density in the QW). In samples with relatively
low Mn content, such as here, the two characteristic temperatures
$T_{c}$ and $T_{CW}$ coincide. Note that, due to the anisotropy of
the heavy hole states, the exchange field is perpendicular to the
QW plane. Thus it corresponds to the normal Faraday configuration
and circular polarizations of emitted light. However, if the
signal is averaged over areas of magnetic domains of both possible
orientations, the observed PL is not circularly polarized
\cite{20}. It was shown \cite{27} that even in the micro-PL
experiment with resolution better than 1~$\mu$m, PL is averaged
over different domains. The zero-field splitting measured in
micro-PL gives us the average value of the local magnetization
within the domains present in the imaged spot.

When the spin splitting exceeds the X$^+$ binding energy, a double
line is observed. The high energy component has the ground state
of the hole gas as a final state, and it shows a clear phonon
replica. The low-energy component leaves the hole gas in an
excited state: the energy of this final state monotonously
increases with $p$ (from about 2.5~meV for vanishing values of
$p$, to the Fermi energy for high enough values of $p$ \cite{26}),
so that the observed splitting is a good measure of the local
carrier density (which was calibrated in Ref.\ \onlinecite{26} for
carrier densities in same the range as in the present study). This
is illustrated in Fig.1b, for two different values of the carrier
density (as determined from the well known Moss- Burstein shift,
i.e., the shift between PL and absorption, which was calibrated in
\cite{26}) in a sample with 1.4\% Mn (so that absorption could be
measured through the Cd$_{0.96}$Zn$_{0.04}$Te substrate). This
splitting does not depend on temperature (as soon as the spin
splitting is large enough), which makes it easy to distinguish
from the spontaneous Zeeman effect. Destabilization of the X$^+$
by the giant Zeeman effect due to spontaneous magnetization was
observed, e.g., at low enough temperature. In the present study,
the spontaneous magnetization was always smaller than the value
needed to destabilize the X$^+$.

Finally, the local value of $x_{eff}$ was deduced from the
field-induced shift between two values of the applied field, 1 and
3~T. Note that for such values of the spin splitting, there is no
doubt that the hole gas is fully polarized, so that
carrier-carrier interactions do not change with field and
temperature. At incomplete spin polarisation, carrier-carrier
interactions leads to additional shift of the PL line
\cite{23,26}. Also, the effect of the exchange field is negligibly
small at such field values.

Figure 2 shows typical maps of the different parameters describing
the PL spectra, obtained with sub-micrometer spatial resolution,
and selected from the ensemble of measurements performed on
different areas of the sample. PL was excited directly in the QW
(at 680 nm), and the excitation power was checked to be low enough
to avoid significant heating of the Mn system. Fig.~2a shows the
zero-field PL splitting at 1.8~K, which is the measure of the
local magnetization within the domains. Fig. 2b shows the
splitting between the two components of the $\sigma^{+}$ PL line
at $B=1$ T, which depends on the local hole density. Other maps
(not shown here) were plotted for the Zeeman shift between 1~T and
3~T, which corresponds to the spin density, and for the total PL
intensity.

\begin{figure}
\includegraphics[width=8.5cm]{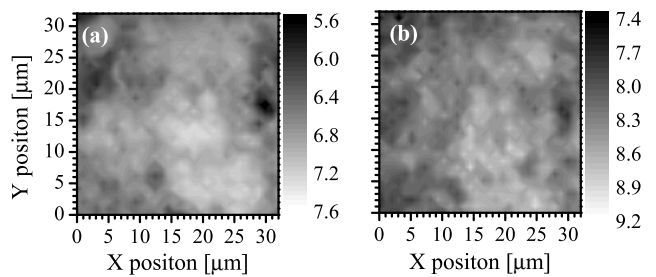}
\caption{Micro-PL maps of (a) zero field splitting (in meV) and
(b) splitting between two components of a $\sigma ^+$ PL line at
$B=1$~T (in meV) }
\end{figure}

All maps (except the map of total intensity) exhibit significant,
irregular, spatial fluctuations. Their characteristic scale is a
few micrometers. These fluctuations of the spectroscopic
parameters reveal fluctuations of the local magnetization, hole
density and Mn content, respectively, for the maps of zero-field
splitting, splitting at $B = 1 \mathrm{T}$, and high-field Zeeman
shift.

In order to decide what is the nature and the origin of the
fluctuations of the spontaneous magnetization, we analyzed the
correlations with the fluctuations of the Mn content (high field
Zeeman shift, fig 3a) and hole density (high-field splitting, fig
3b). We have also checked (not shown) that there is no correlation
between the fluctuations of carrier density and those of spin
density. Only figure~3b evidences a meaningful correlation. On the
1024 points of Fig.~3b, the Pearson's correlation coefficient,
defined as $R = \langle x y \rangle / \sqrt{\langle x^{2}\rangle
\langle y^2 \rangle}$ where $ \langle x^2 \rangle$ is the centered
second moment of $x$, is $\approx $58\%; for such statistics, a
correlation coefficient larger than 8\% ensures a significance
level of 0.01 (i.e., 99\% confidence). This demonstrates that the
main mechanism responsible for the fluctuations of spontaneous
magnetization is the presence of fluctuations of the hole density.
On the other hand, no clear correlation appears in Fig.~3a (on 256
points, $R \approx $ 7\%, much less than the threshold of 16\%,).
Note that this may simply indicate that the distribution of spin
density is too narrow to have any detectable influence on the
spontaneous magnetization. On a macroscopic scale, the spontaneous
magnetization was found to be proportional to both the effective
spin density and the carrier density [19,20]. This dependence is
shown in Fig. 3b by a solid line. It agrees with the correlation
existing between the local values of the two parameters.

\begin{figure}
\includegraphics[width=8.5cm]{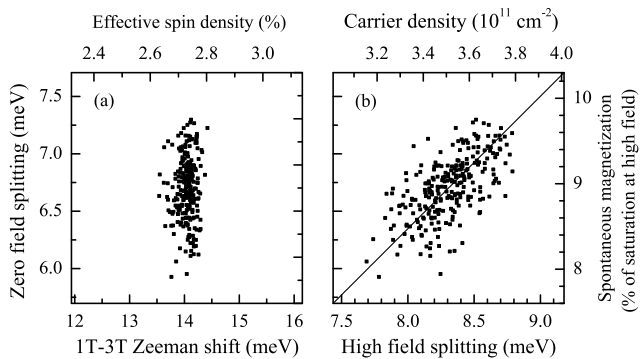}
\caption{Plot of the zero field splitting as a function of (a) the
Zeeman shift between 1 and 3~T, and (b) the splitting measured at
high field, using data from Fig. 2. The three scales have been
chosen so as to show a $\pm 15$\% variation around the mean value.
The top scales are deduced from the measured data using (a)
available parameters of the giant Zeeman effect in CdMnTe
\cite{22} and (b) the calibration of a hole gas in a CdMnTe QW
\cite{26}. The magnetization scale (right scale) uses the giant
Zeeman effect measured at large field on a large area of the
present sample. The solid line shows a linear dependence of the
spontaneous magnetization on carrier density.}
\end{figure}

Such fluctuations of the carrier density might result from
electrostatic fluctuations, or fluctuations of the QW parameters
such as strain and QW width. Fluctuations of the QW width should
have rather similar effects on the carriers and on the excitons.
They are probably small in the case of samples grown at
$280^\circ$C, where pseudo-smooth interfaces are found (i.e., the
characteristic scale is smaller than the exciton coherence length)
\cite{25}. Strain fluctuations are also minimized by a careful
design of the sample structure, so that the whole layer is
coherently strained to the substrate. The main source of
fluctuations of carrier density is thus probably the fluctuations
of electrostatic potential, due to the distribution of ionized
acceptors. One generally considers \cite{28} that the typical
spatial scale of these fluctuations is determined by two
parameters: the distance from the QW to the doping layer (in our
case, with doping from surface states, this is the thickness of
the cap layer, 25 nm), and the inverse of the Fermi wavevector
(about 12~nm for $p = 3 \times 10^{11} \mathrm{cm}^{-2}$).

To conclude, in spite of a rather low spatial resolution (0.5
$\mu$m) inherent to a micro-PL set-up, the fluctuations of the
carrier density in a modulation-doped p-type
Cd$_{0.96}$Mn$_{0.04}$Te QW have been imaged, and their role on
the spontaneous magnetization due to carrier induced magnetism
demonstrated. The effect of spin density fluctuations is too small
to be detected. We may anticipate that the role of electrostatic
fluctuations should be even more dominant in a thick layer, where
the acceptors are present in the DMS layer. Such an optical
determination of fluctuations should be feasible also in (Cd,Mn)Te
QWs with a higher Mn content, where stronger disorder seems to
induce a difference between the Curie-Weiss temperature measured
in the paramagnetic phase under applied field, and the critical
temperature at zero field \cite{29}

Work partially supported by Polish State Committee for Scientific
Research (KBN) grant 5 P03B 023 20.

\end{document}